\documentclass[12pt]{article}
\usepackage[ansinew]{inputenc}
\usepackage{color}
\usepackage{graphicx}
\usepackage{enumerate,latexsym}
\usepackage{amsmath,amssymb}
\usepackage{amsmath,amssymb}
\usepackage{amsxtra}
\usepackage{amsfonts}
\usepackage{enumerate}
\usepackage{cancel}
\usepackage{graphicx}
\usepackage{float}
\usepackage{listings}
\usepackage{indentfirst}

\begin{document}

\title{Near Earth Asteroids:The Celestial Chariots}

\author{Marc Green, Justin Hess, Tom Lacroix, Jordan Marchetto,\\
Erik McCaffrey, Erik Scougal and Mayer Humi \\
Worcester Polytechnic Institute, \\
Worcester, MA 01609 \thanks {e-mail: mhumi@wpi.edu. }}

\maketitle

\begin{abstract}
In this paper we put forward a proposal to use Near Earth Objects as 
radiation shield for deep space exploration. In principle these objects 
can provide also a spacious habitat for the astronauts and their 
supplies on their journeys. We undertake also a detailed assessment of 
this proposal for a mission from Earth to Mars.

\end{abstract}

\newpage
\section{Introduction} 

In the last half-century, humans have set foot on the moon and, as a result, deep space travel seems to be within Humanity's reach. However, there are many issues that must be resolved before space travel to other objects in the solar system can become a reality. Examples of these issues include radiation in space, physiological effects of zero gravity, power generation, and propulsion methods.

In this paper we put forward a novel proposal that can mitigate the dangers that radiation can pose to space travel. The basic idea behind this proposal is the use of Near Earth Objects (NEOs) as a temporary shelter from radiation for astronauts on deep space missions, like a voyage to Mars.

Just as fire can be both dangerous and beneficial, the NEOs that pose a danger to Humanity on Earth can be used as "Celestial chariots." By protecting the astronauts from radiation, and providing spacious housing and storage facilities, long journeys into deep space become possible.

We begin with a general overview of radiation in space and its effects on exposed humans. Various strategies to protect astronauts from these harmful effects are then described. We concentrate on the use of NEOs a radiation shields, and we put forth a proposal using them on a trip to Mars and the feasibility of such a proposal. We conclude by discussing ideas that, if realized, would greatly increase the likelihood and the benefit of NEOs as radiation shields.

To examine this idea in detail, one has to choose the most suitable NEO out of the ten thousand known that could be utilized to travel between Earth and Mars. The chosen NEO must be of sufficient size and material to properly shield its cargo of humans and supplies from the radiation in space. In addition, we examine the possibility of "domesticating" an asteroid by manipulating its orbit. To protect the astronauts on their journey, a habitat can be built in the NEO in which a manned crew would ride safely from Earth to Mars. The crew would then exit the NEO upon approaching Mars. 

\section{Radiation and Its Effects}

Short wavelength radiation, such as X-rays and Gamma rays, is harmful to living organisms. On Earth, protection is provided from these harmful effects by the Van-Allen belts, Earth's magnetic field, and the atmosphere. In space, however, there is no such shielding.

Radiation in space is made up of Galactic Cosmic Rays (GCRs) and Solar Particle Events (SPEs), which are atomic and subatomic particles accelerated to (very) high energy levels. Another source of radiation, though usually ignored, is known as secondary particles [1]. These high-energy radiation sources pose a major threat to astronauts as they can cause cancer and genetic mutations. Humans outside the Earth's protective shield must use other means to protect themselves from radiation's ill effects.

\subsection{Galactic Cosmic Rays} 

GCRs originate from outside our solar system and have particle energies known to peak around 1 GeV within the solar system. Approximately 98 percent of these particles are protons or heavy ions which possess a high linear energy. This linear energy enables these particles to deeply penetrate almost any shielding. The collision of the radiation with a shielding material is likely to generate secondary particles, increasing the total amount of shielding needed to maintain a safe environment for life.

GCR radiation levels within the solar system vary with the natural solar cycle.  The fluctuation of the Sun's magnetosphere and material output changes the amount of GCRs deflected away from the core of the solar system. [2]

\subsection{Solar Particle Events}

SPEs are made of solar flares and coronal mass ejections (CMEs), both of which spew ionized gases and radiation from the Sun. Large events of this nature are rare, however their frequency depends upon the sun's 11-year activity cycle.

Impulsive flares are short lived, on the order of hours, but release large amounts of electron radiation. They are not overly dangerous to space travel to other planets as they are emitted in a range from 30 to 45 angles in solar longitude and leave the ecliptic plane quickly.

CMEs are much longer lived than flares, lasting days, and are characterized by a high proton flux on the order of $10^9$ particles per square centimeter. They can erupt anywhere from the Sun's surface and reach Earth and other planets, impacting spacecraft and space-based resources en route to or in orbit around these planets.[3]

\subsection{Impact of Radiation on Living Organisms} 

Radiation in the International System of Units (SI) can be measured in Sieverts (Sv). The Sievert is the equivalent absorption of one joule of energy by one kilogram of matter. This means that it represents the biological effects of ionizing radiation by taking into account the biological context.

A human exposed to more than 1 Sv will suffer varying illnesses including leukopenia and other immune system-impairing conditions. A radiation dose greater than 8 Sv will be fatal to humans within days, as they succumb to organ failure and severe burns. An astronaut on board a spacecraft exposed to a solar flare or CME would likely be killed.

Even on spacecrafts without humans, radiation can pose a hazard to the mission. Space radiation can interact with the spacecraft electronics, damaging or destroying them. It affects computer operations and stored data by flipping the charge of bits between positive and negative. This happens when a charged particle impacts a bit, it flipping the charge and corrupting the data. Under extreme radiation, like the particle flux of CMEs, the information can be completely corrupted.

Currently, there are two main ways to internally shield electronics. By using older chip designs with larger gaps between bits, it is less likely that a particle will flip multiple bits, minimizing the damage. Additionally, the CPU equivalent of lightning rods can be built into the vulnerable circuitry to redirect the incoming radiation into less-critical areas. An external method to shield the electronics involves physically encasing the computers with protective material, but at the cost of adding mass to the mission. It is desirable to find the lightest and most cost effective shielding.

\subsection{Radiation Shielding}

Radiation shielding is measured in "half thickness", which is the amount of material needed to reduce the incoming radiation in half. By overlapping several half thicknesses, incoming radiation can be further reduced and a safe environment can be established. If we know the expected radiation levels and the material properties of our shield, we can determine how much shielding is needed. The following equation can be used to determine the half thickness, where $\mu$ is the material's bulk mass absorption coefficient and $\rho$ is the material density.

$$ 
t_{1/2}=-\frac{\ln(0.5)}{\mu\rho}
$$ 

Determining the shielding thickness needed is simple in concept, but requires knowledge of High Energy Particle Physics and Material Science. The interaction between the incoming radiation with the shielding materials needs to be accurately known.

To keep the astronauts healthy throughout their travel in space, it is desirable to keep the radiation levels in the spacecraft as close to Earth-level as possible. There are established guidelines published by the NCRP (National Council on Radiation Protection) which NASA currently uses to determine what the astronauts on the ISS can safely experience.[4]

\subsection{Radiation Blocking Materials}
 
Shielding materials should be selected based on several criteria including half thickness and total mass. The ideal material would be both lightweight and have good deflection properties against radiation particles. Lead is the de-facto shielding material on Earth, but it does not meet our criteria due to it's large mass. Even though it will not be used for spacecraft shielding, it will be used as a basis for comparison with the other materials because it is a common shielding material.[4] Below, in Table 1, we provide a list of materials that could be used to protect against radiation in future space missions. In this table we explain what kind of material it is, how it can be used against radiation, and how much of the material is required. 

Table 1: A Sample of Radiation Protective Materials
{\small
$$\begin{array}{|c| c| c| c|} \hline
& & & \\ [-2ex]
\textrm{\tiny Protective Material} & \textrm{\tiny What is it} &
\textrm{\tiny How does it work}& \textrm{\tiny How much is required}\\
& & \textrm{\tiny against radiation} & \\
\hline
\textrm{\tiny Lead} &  \textrm{\tiny Chemical element with atomic number} & \textrm{\tiny It's extremely high density provides}
& \textrm{\tiny 10cm for a reduction of ~1000x}\\
 & \textrm{\tiny 82} &  \textrm{\tiny shielding from radiation particles} & \\
\hline
\textrm{\tiny Polyethylene} & \textrm{\tiny Most commonly used plastic for} &
\textrm{\tiny Demron is lightweight, flexible and}&
\textrm{\tiny A thickness of 2.7 cm (72 layers) and 29}\\
\textrm{\tiny(Demron)} & \textrm{\tiny commercial products} &
\textrm{\tiny contains proprietary materials that} &
\textrm{\tiny cm (240 layers) of Demron would be}\\
&\textrm{\tiny A chemically synthesized polymer} & 
\textrm{\tiny block radiation.} \textrm{\tiny It can be treated like} &
\textrm{\tiny required for a two factor and ten factor }\\
& \textrm{\tiny with high amounts of hydrogen} & 
\textrm{\tiny a fabric for cleaning, storage, and } &
\textrm{\tiny reduction in transmission} \\
&\textrm{\tiny -Demron is polyethylene between two} & 
\textrm{\tiny disposal purposes} & \\
& \textrm{\tiny layers of fabric} & & \\
\hline
\textrm{\tiny Boron Nitride} &\textrm{\tiny An equal chemical combination of}
&\textrm{\tiny Due to its light nucleus, can}
&\textrm{\tiny Approximately 1.5m is needed to reduce} \\
\textrm{\tiny Nanotubes} & \textrm{\tiny both Boron and Nitrogen nanotube}
&\textrm{\tiny successfully absorb harmful neutrons} &
\textrm{\tiny the effective dose rate (Eiso) by 45-} \\
&\textrm{\tiny containing Boron Nitride} & 
\textrm{\tiny in secondary radiation. Additionally} & \textrm{\tiny 48\%}\\
& & \textrm{\tiny if used in the development of a space} & \\
& &\textrm{\tiny shuttle, this can further decrease the} & \\
& & \textrm{\tiny harm of radiation exposure for} & \\
& & \textrm{\tiny astronauts.} & \\
\hline
\textrm{\tiny Electrostatic} &
\textrm{\tiny A material capable of blocking the}& 
\textrm{\tiny The use of conductive materials such} &
\textrm{\tiny Several feet of carbon a nanotube are} \\
\textrm{\tiny Shielding \& Carbon} & 
\textrm{\tiny effects of an electric field, while also} &
\textrm{\tiny as  carbon-nanotubes (CNTs) can} &
\textrm{\tiny required to shield against radiation} \\
\textrm{\tiny Nanotubes} &
\textrm{\tiny allowing the passage to magnetic} &
\textrm{\tiny conduct enough energy to generate} & \\
& \textrm{\tiny fields.} &
\textrm{\tiny an electrostatic shield capable of} & \\
& & \textrm{\tiny blocking all incoming ion particles} & \\
\hline
\textrm{\tiny C60 (Buckminster-} &
\textrm{\tiny A spherical fullerene molecule with} &
\textrm{\tiny Provides potential benefits against} &
\textrm{\tiny TBD, more research needs to be done to}\\
\textrm{\tiny fullerene)} &
\textrm{\tiny the formula C60 with a cage-like fused} &
\textrm{\tiny radiation if used as an antioxidant}&
\textrm{\tiny determine appropriate doses} \\
&\textrm{\tiny ring structure} &
\textrm{\tiny drug} & \\
\hline
\end{array}$$
}
\section{Near Earth Objects} 

Near Earth Objects (NEOs) are defined as any object that passes within 0.3 AU of Earth at some point in their orbit around the Sun. These include, but are not limited to, asteroids and dead comets. NEOs pose a danger to humanity on Earth as even the smallest NEO impact can release as much energy as a nuclear weapon or volcanic eruption. The Chelyabinsk Meteor that struck Russia on February 15th, 2013, was approximately 18 meters in diameter and generated nearly 440 kilotons of TNT. The threat of NEOs to humanity is very real, but what if they could be tamed and used for space travel?

\subsection{NEOs for Radiation Protection And Space Travel}

We propose to use NEOs for transportation within the solar system. With proper asteroid selection and preparation, it will be possible to take advantage of their material properties to protect the astronauts from space radiation. Other benefits of using NEOs will be spacious living quarters and expanded storage facilities for food, water and medical and scientific supplies. Larger crews can thus be supported, which will allow for social interaction during the long voyage and help the psychological well-being of the crew. For the purposes of explaining our proposal, we shall use a mission to Mars as a case study.
 
\subsection{The basic Mission Profile} 
\begin{enumerate}
\item Identify NEOs that meet orbital requirements to pass Mars and Earth. 
\item Determine NEO composition through physical landing, practicing orbital 
maneuvers.
\item Prepare NEO for habitation through robotic missions.
\item Land a crew on board NEO when habitation is completed and the 
asteroid is passing Earth en-route to Mars. 
\item Travel to Mars inside NEO and leave when at Mars. 
\end{enumerate}

The implementation of this program is not simple. Although we have the technology to precisely calculate the orbits of NEOs, at present, it is quite difficult to determine if a particular NEO will be suitable as a radiation shield.

Furthermore, efficient methods have to be developed to prepare an asteroid for habitation. Below, we propose some ideas to reduce these complications, including changing the trajectory of asteroids. While these are all obstacles to the success of our vision, we believe that they are not insurmountable.

These preparations will need to be done ahead of time through robotic missions. Likely, a single launch or quick series of launches will transfer all the materials needed for the construction. Until the asteroid passes close to Earth again, the robots will build the habitable and radiation shielded volume. This allows for astronauts to simply land on the asteroid and have everything prepared. Additionally, it will be needed to plan for the return trip.

\subsection{Advantages of Using NEOs} 

One of the consequences of using a conventionally shielded spacecraft is the large amount of mass that must be put into space. The cost of doing this behooves mission planners to design the interplanetary spacecraft with longevity in mind to reduce the need for expensive launches. However, long-term operation in high-radiation environments is dangerous, as radiation embrittles metals over time. This can best be seen in nuclear submarine reactors, where thirty years of service takes its toll on the reactor walls. Metal embrittlement and fatigue are a major concern for conventional spacecrafts where the interior has to be pressurized. However, this issue becomes moot if NEOs are used. While aluminum and other materials might be needed to construct the living quarters, they will be shielded by the bulk of the asteroid, and not weaken from radiation. Additionally, building inside the NEO allows for pressurization stresses to be spread into the NEO, reducing the stresses in the construction material.  

Furthermore, one of the most dangerous physical threats to spacecraft, micro impacts, can be all but eliminated using NEOs. Small, high velocity debris generated in collisions can rupture conventional spacecraft. They are extremely hard to detect due to their size, but can vent the spacecraft's atmosphere into space, killing its crew. Nestled deep inside an NEO, micro impacts will not be able to penetrate far enough to be a threat to the crew.

\subsection{Planning a Mars Mission} 

There is a lot to consider when constructing a mission to Mars using NEOs as transportation and radiation protection. Due to the large amount of unknown data, we cannot be certain that a given solution will work. However, we believe that the solution we have provided is flexible enough to accommodate the unexpected.

\subsubsection{Enumeration of Asteroids} 

Selecting the right NEO out of the tens of thousands in orbit around the Sun is a daunting challenge. We shall limit the search to asteroids, as they are more likely to provide the radiation shielding we require compared to the other objects. From the more than 9,500 asteroids presently identified by NASA and other Space Agencies, our asteroid will have to meet the following criteria.

\begin{enumerate}

\item Must approach Earth and Mars' orbits and meet other requirements to 
improve crew transfer and safety. 
\item Must be large enough and of the right composition to provide adequate 
shielding. 
\item Must be small enough to be sufficiently movable to fine-tune the orbit. 
Our target size for such an asteroid is approximately 100 meters in diameter. 
\end{enumerate}

Our target size for such an asteroid is approximately 100 meters in diameter. This offers us the highest likelihood that we will have enough shielding, living space, and it will be small enough to move. Unfortunately, current detection methods are not accurate enough to detect asteroids 100 meters in diameter of less, and future satellite asteroid observation missions should be conducted to increase the likelihood of finding ideal asteroids. Infrared telescopes would be ideally suited for our purposes as they can sometimes identify the composition of the asteroid using spectrology.

\subsubsection{Orbital Criteria} 

Even if we find an asteroid with a perfect material composition for radiation shielding, it means nothing if that asteroid does not fly by Earth and Mars on a regular basis. Therefore orbital criteria are an essential part of selecting asteroids. To reduce the amount of orbital manipulation required for each potential asteroid, those closest to Earth, which also intercept Mars' orbit, will be examined. Their perihelion should be within 0.05AU of Earth's orbit and their aphelion should extend to between Mars perihelion and aphelion, a range of 1.4 to 1.7 AU. The aphelion of their orbits should not extend beyond 1.7 AU to avoid a collision with another asteroid in the Asteroid Belt, thus drastically altering their orbit and making them useless for our purpose. These parameters reduce the number of Near Earth Object asteroids from over 9500 to around 200.

A final orbital consideration is the inclination of the asteroids orbits relative to Earth and Mars. In an Ecliptic plane reference frame, which passes through the Sun and Earth orbit, Mars orbit has an inclination of -1.8 degrees. In order for Asteroids to approach both Earth and Mars, they must orbit in between the planets inclination angle. For simplicity, we shall assume that we can manipulate the inclination angle of those asteroids nearest to the ideal orbit in such a way that they will precisely intercept the planets. By limiting our search to +/-10 degrees from the Ecliptic plane, the number of potential asteroids is
refined to 43 (these are the highlighted asteroids in the Appendices). 
A complete list of known asteroids and their orbital parameters is available in Ref $17$. Unfortunately, not much, if anything is known about their physical properties and currently we can only identify potential asteroids by their orbit.

An artistic conceptual trip to Mars using NEOs is attached at the end of
this paper

\subsubsection{Asteroids Composition} 

Asteroids that are classified as NEOs are broken into several categories based on their orbital path and material composition. One particular class of NEOs, known as Chondrites, are made of iron, water and carbon. Based upon infrared spectrology, NASA estimates that these asteroids are approximately 88 percent iron, making them an ideal radiation shields from a material standpoint. However, since iron is difficult to dig into, other asteroids that have a looser, or less-dense composition might be more suitable. With reduced density, they would have to be bigger to account for the difference in half thickness, but in space size doesn't matter. If the asteroid's mass is lower it will require less work to change its orbit.

Iron is the second best metallic element used to protect against gamma radiation being only second to lead. It only takes 4 inches of Iron to reduce gamma radiation damage by a factor of 10, whereas it takes 24 inches of water to reduce the damage by the same amount. Drilling down 340 feet of iron would reduce radiation damage by a factor of 1 million. The optimal choice of an NEO for mission purposes will need to take into account other properties of NEOs, such as their temperatures.

\subsubsection{The Unknowns} 

While an asteroid may fit the orbital requirements for use in a manned mission, if it is incapable of shielding astronauts from radiation, it is useless. Not much is known about the material and physical makeup of the vast majority of asteroids. Therefore the necessary shielding thickness cannot be computed to construct safe living spaces on board. Experiments performed on samples of each asteroid, can provide the necessary data.

To precisely identify how much shielding is needed, samples of asteroids need to be collected and analyze. Using instruments similar to the Curiosity Rover x-ray spectrometer and Laser Induced Breakdown Spectroscopy system, a probe exploring each potential asteroid can relay the exact material composition and makeup to Earth.

Attempting to rendezvous with the most promising asteroids for sample analysis is not a new idea. In 2000, Japan launched the Hayabusa mission which landed on 25143 Itokawa and was able to retrieve samples and return them to Earth.[7] Complications in the attempts to collect the samples leads us to believe that a spacecraft designed to stay with the asteroid and analyse it would be better. It would allow for more information to be collected about the asteroid and possibly begin mapping the surface aid in construction.

\section{Futuristic Vistas} 

\subsection{Orbit Modification} 

Making feasible use of asteroids for transportation will often require that the asteroid's current orbit be modified in order to make the transfer between Earth, Mars, and the asteroid more convenient. Unfortunately, this process requires extensive use of impulsive thrusting over the course of several orbits. The thrusting will have to be done over perigee and apogee centered burn-arcs, possibly at constantly changing thrust angles. Furthermore, concurrent changes in the asteroids apogee, perigee, eccentricity, and inclination angle will have to be made in order to affect the orbit change in a reasonable amount of time and minimum amount of resources.[8] Due to the difficulty in analyzing such a scenario the specifics will not be covered in this paper. However the nature of the propulsion that should be used in this process can be analyzed. The basis for analyzing different propulsion methods lies in an equation of motion derived from the Reynolds transport theorem. This equation is:

$$ 
\left(M-\frac{dm}{dt}\right)A=\frac{dm}{dt}V
$$ 

Where: 
\begin{itemize}
\item M is the total mass of the craft (the asteroid)
\item A is the acceleration of the craft 
\item t is thrust time 
\item $\frac{dm}{dt}$ is the mass flow rate of the propulsion system. 
\item V is the propellant velocity of the propulsion system 
\end{itemize}

The focus will be on the right half of the equation, $\frac{dm}{dt}V$, which is equal in 
magnitude to the thrust force on the spacecraft. These two parameters, 
the mass flow rate $\frac{dm}{dt}$, and the propellant velocity (V), can be used to 
evaluate the advantages and disadvantages of propulsion methods. The 
advantage of a high V is that more energy is imparted per propellant mass,
meaning that in a situation where propellant mass is limited, such as long 
distance missions, more total energy can be imparted. High V systems tend to 
have very low $\frac{dm}{dt}$, but also very low thrust. However many high thrust propulsion technologies have low V, making them mass inefficient and suitable 
only for short missions. 

Since propellant will be brought to the asteroid very infrequently due to the rarity of close approaches and the difficulty of coordinating long distance 
shipments, the propulsion method used on the asteroid will have to be as 
propellant mass efficient as possible. To this end a high propellant velocity is necessary, which can be found in ion engines. 

An ion engine generates thrust by accelerating ions to high velocity and 
expelling them out of the rear of the spacecraft. The gas is typically an inert 
gas such as Xenon to avoid unwanted reactions. [10] There are two ways to accelerate the gas. First is the electron bombardment method where the 
engine bombards the propellant with high-energy electrons, knocking Xenon's 
valence electrons free. [11] Second is the electron cyclotron resonance method, which excites the electrons in the gas via microwaves and magnetic 
fields. [11] Once ionized, the gas is accelerated out of the engine by 
electrostatic forces generated by a positively charged grid at the beginning of the flow chamber and a negatively charged grid at the end. [12] 

Some high velocity ion engines expel their propellant at speeds in excess of 
90,000 m/s, but produce minimal thrust and expel very little propellant. [13] Atypical modern ion engine thrust is a mere 0.5 N, but fortunately higher thrust 
engines are in development.[16] One such project, the High Power Electric 
Propulsion project has developed a 40 kW engine, more than 19 times more powerful than the engine used by Deep Space. [17] [18]. Furthermore, 200 kW 
configurations have been suggested, which would boost thrusts up to 18 N, 
propellant velocities of 40,000 m/s or higher, and with a low mass flow rate between 100 and 1200 mg/s.[19] Despite the power increase, they retain a 
power efficiency of upwards of $60\%$ and a calculated thrust to mass flow ratio 
of up to 50000 N*s/kg. 

If several of these higher thrust ion engines are placed on the asteroid they 
will allow for the asteroid's orbit to be modified with low propellant and energy costs. The main objective of these orbit manipulations will be to place the 
asteroid into a mildly elliptical orbit which will travel from Earth to Mars in 
about 200-250 days. This will require the asteroid's velocity to be approximately 25 kilometers per second, an easily attainable figure. If the 
engines are thrusting for years on end. 

\subsection{NEO Domestication} 

The solution to the radiation problem that we are proposing in this paper relies on having a collection of easily accessible domesticated NEOs, which we have 
implicitly assumed to exist in our presentation. However, this is where the 
complexity of our solution lies, and this is what will determine the likeliness of using NEOs as transportation vehicles within the solar system. 

Domesticating a single NEO involves modifying the NEO orbit and excavating a
radiation-protected area in its body. As previously discussed, it is plausible to 
modify an NEO orbit with ion thrusters. Further research will need to be done 
in order to determine the best methods to create a radiation-protected habitat,
but we can put forth here the essential requirements. Using advanced robots 
that are physically and electronically hardened to survive in deep space, the 
selected asteroid will be prepared by drilling away material to construct the 
shielded living quarters for the crew. Being essentially stranded on the asteroid with no material support and little communication, the robots will 
have to be rugged, capable of repairing themselves, and completing the 
complex operations demanded of them with minimal error. 

The vision we have is not just one domesticated asteroid. We believe that a 
few dozens will be necessary in order to make frequent trips to and from Mars 
due to their long orbital periods. Great care should be taken when creating this group so that the asteroids will not be on a collision course with Earth and 
survive for centuries to make the most of our investment in converting them. 

\subsection{Time of Flight} 

When dealing with radiation and space travel, time is in the essence. 
It dictates 
how much shielding, food, water and fuel is needed for the mission, and 
therefore something that we should be aware of during mission planning.
Trying to calculate the close approaches between the asteroid and each planet 
is difficult when the different orbit periods and angle differences are 
considered. In some cases it might be necessary for the crew to stay 
on board the asteroid for several years while the planets move into position. On the 
other hand, if everything went according to plan, an asteroid could take a 
direct route between Earth and Mars. 

With orbit modification, this direct route could become more routinely possible 
if enough asteroids were domesticated and enough orbit modifications done. 
In this manner, the travel times between the planets would be around 200-400
days. Most of the asteroids that met our orbital criteria typically orbited the 
Sun in 500 to 650 days and they intersect each orbit in two places, creating 
two transfer points from planet to asteroid or vise-versa. 

Getting to and from the asteroid is another concern for astronauts, as their 
exposure to radiation is greatest during this transfer. Given that the asteroid 
cannot come into Earth's gravitational influence, the astronauts must exceed 
Earth's escape velocity of 11920 m/s. From launch, it should only take a few 
days to leave Earth's gravity well and rendezvous with the asteroid. The total 
time of flight for this portion of the mission is solely dependent on how close the asteroid is to Earth. With precise orbit modifications the asteroids can be 
put a safe distance from Earth's gravity, but not too far away that it risks the 
health of the astronauts. Leaving from Mars should be very similar to 
this process, but quicker as Mars gravity is weaker than Earth's 

Leaving the asteroid will be different than landing on it, as the escape velocities are thousands of 
times lower than any planet. In theory this transfer should be a simple reverse of leaving a planet, 
but this time, the astronauts have to enter the atmosphere and precisely land at their target. With 
proper care and planning, this can be done, but it will not be simple. 

\subsection{Returns on Investment }

Domesticating a fleet of asteroids will not be cheap, nor perhaps cost effective in the short term, but it is the long-range returns that make NEO transportation worthwhile. The international Space Station has orbited for more than 15 years and will continue doing so for many years to come, but at some point the space station will reach the end of it's life and be retired. The data gathered onboard the ISS cannot be collected anywhere else, making it well worth the 150 billion dollars it has cost so far. Thus, the domesticated asteroid orbiting around the Sun will be worth the trillions invested in them as they serve humanity. The benefits of such a mission extend much further beyond the simple monetary, scientific, and humanitarian returns it will generate. It offers a way to protect Earth and humanity. To date, Earth has no defense against asteroids and comets and we bear the scars of massive impacts in the past. The likelihood of such a collision is relatively admittedly small, but using the orbit modification techniques developed to domesticate the asteroids, humans can remove the most dangerous NEOs that threaten Earth and humanity. The safety of our planet and continuation of our species is worth any price.

\section{Conclusion} 

We believe humanity is on verge of new era in which humans will expand their habitat to other celestial bodies. The advent of this new era is driven by the combination of human curiosity and drive to einsure human survival. Radiation poses one of the greatest threats to the successful expansion of our race throughout the solar system and beyond. Without proper protection from this hazard, future space travelers may either die or be genetically altered. This paper put forward the idea that the use of NEOs is a possible solution for protecting humans against radiation en route to a new planet or celestial object. This idea should be investigated further in the future as a supplement to other radiation protection efforts.

Although NEOs have primarily been viewed as hazardous to Human survival on Earth, we have developed a different perspective allowing us to take advantage of the properties that make them dangerous. In the same way that fire, an incredibly destructive force, was mastered in order to advance and expand human society on Earth, we believe that NEOs can be used are needed to expand human society across space.

\section*{References}

\begin{enumerate}

\item Adamczyk, A. (2013, January 23). Radiation environment during space 
flight and on other planets. \\
http://hps.org/publicinformation/ate/faqs/spaceradiation.html

\item Brown, D. L., Beal, B. E., \& Haas, J. M. (2009, October 27). 
Air Force Research Laboratory High Power Electric Propulsion Technology
Development, \\
http://ieeexplore.ieee.org/stamp/stamp.jsp?tp= \&arnumber=5447035\&tag=1

\item Chamberlin, A. (n.d.). Neo discovery statistics. \\
 http://neo.jpl.nasa.gov/stats/

\item Cucinotta, Dr. Francis. Space Rad Health. (April 2001).\\ 
http://www.nasa.gov/centers/johnson/pdf/513049main\_V1-2.pdf

\item Deep Space-1. (2009, April 21). NASA Glenn Research Center \\
 http://www.grc.nasa.gov/WWW/ion/past/90s/ds1.htm

\item Dawn. (2011, July 12). NASA \\
http://www.nasa.gov/mission\_pages/dawn/spacecraft/

\item Electrostatic shielding. 2012 Dictionary of Engineering.\\ 
http://www.dictionaryofengineering.com/definition/electrostatic-shielding.html

\item European Space Agency. (n.d.). Near earth objects - dynamic site. \\ 
http://newton.dm.unipi.it/neodys/index.php

\item Friedman, H. W., \& Singh, M. S. U.S. Department of Energy, (2003). 
Radiation transmission measurements for demron fabric.

\item Goebel, D. M., \& Katz, I. (n.d.). Thruster Principles. 
In D. M. Goebel, \& I. Katz, Fundamentals of Electric Propulsion: 
Ion and Hall Thrusters. Hoboken: John Wiley \& Sons, Inc.

\item Ion Propulsion. (2008, May 21). NASA Glenn Research Center \\
http://www.nasa.gov/centers/glenn/about/fs21grc.html

\item J. E. POLLARD, P. S. (n.d.). Simplified Analysis of Low-Thrust 
Orbital Maneuvers.\\
 http://www.dtic.mil/cgi-bin/GetTRDoc?AD=ADA384536

\item Keith, J. E. (2006). Gas Dynamics. Pearson Education Inc.

\item Matloff, G., \& Wigla, M. (2010). NEOs as stepping stones to mars 
and main-belt asteroids. Acta Astronautica, 68(5-6), 599-602. \\
http://dx.doi.org/10.1016/j.actaastro.2010.02.026

\item McMillan, Robert - Martian Computing is Light on RAM, Heavy on 
Radiation Shielding.. Wired.com (2012) \\
http://www.wired.com/wiredenterprise/2012/08/martian-computing-is-light-on-ram-heavy-on-radiation-shielding/

\item Minkel, J. (2010, July 22). Astronomers find largest molecules 
ever known in space. \\
http://www.space.com/8804-astronomers-find-largest-molecules-space.html

\item NASA JPL Small Body Database. \\
http://ssd.jpl.nasa.gov/sbdb.cgi

\item Understanding space radiation, National Aeronautics and Space 
Administration , Lyndon B. Johnson Space Center. (2002).(FS-2002-10-080-JSC).\\ 
http://srhp.jsc.nasa.gov/

\item Thomas publishing company llc. (2013, February 22). \\
http://www.thomasnet.com/articles/custom-manufacturing-fabricating/radiation-shielding-materials
\end{enumerate}

\newpage
\begin{figure}[ht!]
\centerline{\includegraphics[height=120mm,width=160mm,clip,keepaspectratio]{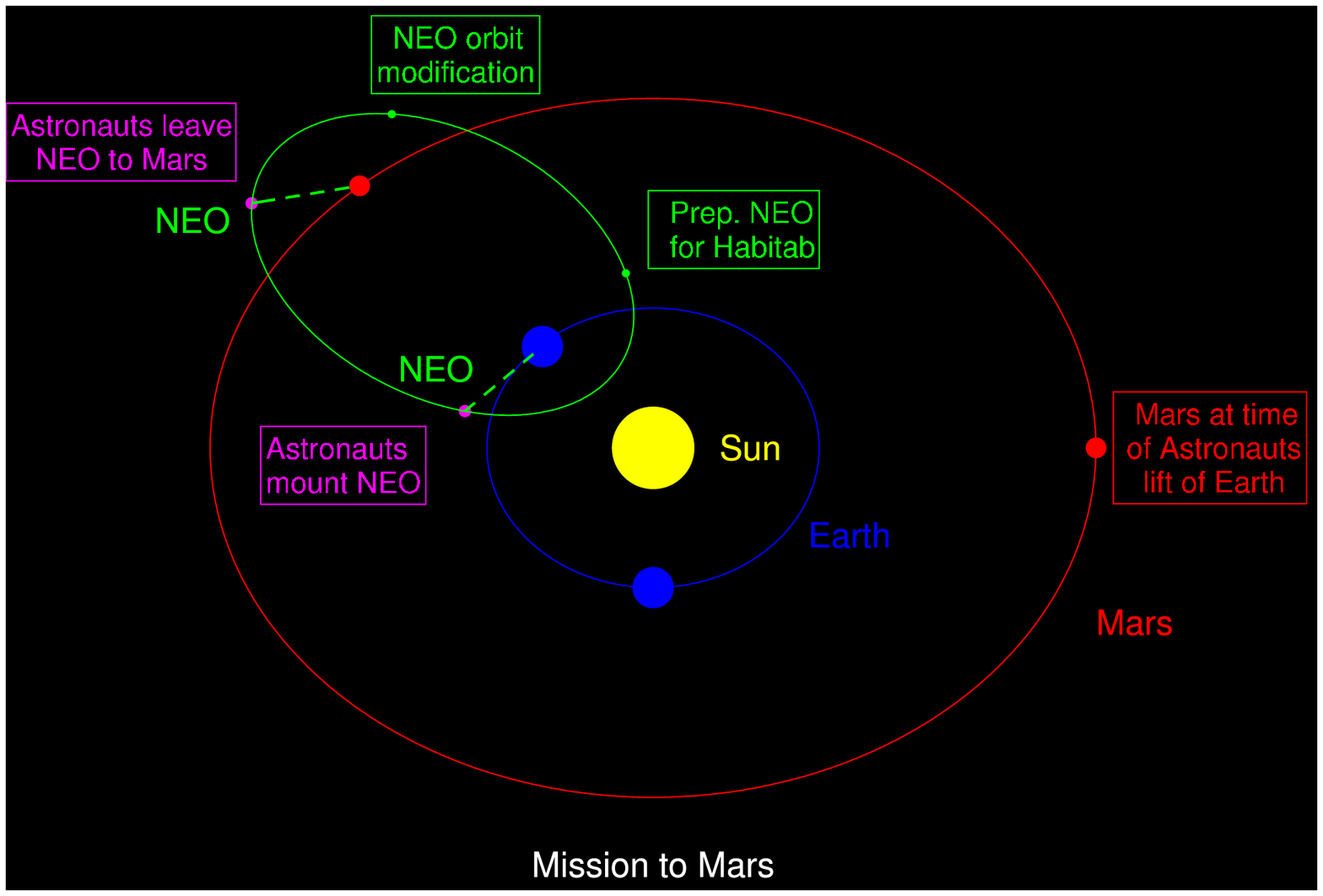}}
\label{Figure 1}
\caption{}
\end{figure}

\end{document}